\documentclass{revtex4}
\usepackage{psfig}
\usepackage{subfigure}
\usepackage{graphicx}
\usepackage{doublespace}
\usepackage{amsmath}

\newcommand{\gtsimeq}{\raisebox{-0.6ex}{$\,\stackrel
        {\raisebox{-.2ex}{$\textstyle >$}}{\sim}\,$}}

\begin{document}
\doublespace

\author{Francesco Sorrentino}
\affiliation{Universit{\`a} degli Studi di Napoli Parthenope, 80143 Napoli, Italy. \\ Institute for Research in Electronics and Applied Physics, University of Maryland, College Park, Maryland 20742, USA. }

\begin{abstract}
In this paper we present an approach in which synchronization of chaos is used to address identification problems
. In particular, 
we are able to identify:
(i) the discontinuity points of systems described by piecewise dynamical equations and (ii) the delays of systems described by delay differential equations.
Delays and discontinuities are widespread features of the dynamics of both natural and manmade systems. The foremost goal of the paper is to present a general and flexible methodology that can be used in a broad variety of identification problems.
\end{abstract}

\title{Identification of delays and discontinuity points of unknown systems \\ by using synchronization of chaos}

\maketitle

Recent work \cite{Abarbanel,Yu:Parlitz,IDTOUT} has shown that synchronization of chaos can be conveniently used as a tool to identifying the dynamical equations of unknown systems \footnote{The idea of using synchronization or control for parameter and model identification was originally presented in \cite{So:Ott:Day}.}. For instance, in \cite{IDTOUT}  a largely unknown chaotic (nonlinear) system was coupled to a model system and a  general adaptive strategy was proposed to make them synchronize by adaptively varying the parameters of the model until they converge on those of the true system. The strategy takes advantage of the fact that complete synchronization of chaos is only possible if the coupled systems are exactly the same \footnote{In order to estimate all the parameters of interest, the \emph{linear independence condition}, presented in Ref. \cite{Yu:Parlitz}, needs to be satisfied.}. 

In linear systems, an \emph{observer} is a model dynamical system that is able to reconstruct the state of an unknown true system from knowledge of its dynamical inputs and outputs.  References \cite{Nij} have outlined the connection between the problem of synchronization of dynamical systems and the problem of the design of an observer to reconstruct the state of an unknown system. The model system introduced in \cite{IDTOUT} acts as an observer that dynamically identifies the parameters and the initial conditions of the nonlinear functions describing the dynamics of the true system, when they belong to a certain class 
(e.g., they are smooth and polynomial up to a given degree).   In general, a condition that needs to be satisfied in order for the observer to fully reconstruct the state of the unknown system is that the inputs are persistently exciting signals, i.e, they solicit all the dynamical modes of the true system. In the case of  chaotic autonomous systems, this requirement is naturally satisfied by the  dynamics, which is in a state of persistent excitation, even in the absence of inputs \footnote{This applies to the relevant dynamical state variables, i.e., to those that evolve on the time scale of chaos. Eventual other variables that evolve on a longer time scale can be considered constant with respect to the faster chaotic dynamics, and as such, regarded as parameters of the model.}. This has important consequences since it allows to possibly extract a large number of unknown quantities from one chaotic time trace, without the need of introducing external inputs (see e.g. \cite{IDTOUT,SOTT2}). On the other hand, a limitation is that stability of synchronization of chaotic systems is usually evaluated locally about the synchronized evolution and for the parameters of the systems being the same (i.e., it relies on the successfulness of the identification strategy).  Therefore, the effectiveness of the strategy 
may be sensitive to the choice of the initial conditions  and it may depend on a careful selection of the adaptive strategy parameters.

Chaos can also arise in systems whose dynamics is described by delay dynamical equations or by  piecewise linear/nonlinear equations. These are both very common situations in nature and in applications. For example, piecewise dynamical equations appear in mechanical systems with impacts \cite{BrogliatoBook}, relay-feedback systems \cite{Relay_PWS}, DC/DC converters \cite{BanerjeeBOOK}. Delay dynamical equations are usually invoked to describe physiological processes in hematology, cardiology, neurology, and psychiatry \cite{Mackey_Glass}; but find application also in chemistry, engineering and technological systems \cite{DDE_BOOK}; examples are lasers subject to optical feedback \cite{DDE_lasers}, high-speed machining \cite{DDE_machining},  mechanical vibrations \cite{DDE_vibrations}, control engineering \cite{Minorski}, and traffic flow models \cite{DDE_traffic}. 
Our goal is introducing a flexible adaptive strategy that can be used to identify either discontinuity points or delays of the equations of unknown chaotic systems from knowledge of their state evolution. To our knowledge, these two important and interesting problems have not been addressed yet in the literature. More in general, the paper presents a general methodology that can be used to solve a broad variety of identification problems, including apparently difficult ones. 



As a first problem of interest, we consider that the unknown true system evolves in discrete time and is described by a piecewise dynamical equation. In particular,  we are concerned with the case that the unknown system evolution obeys,
\begin{equation}
{x}^{k+1}=f_{\sigma}({{x}}^k), \label{dtrue}
\end{equation}
where ${x}^k \in R$, $f_{\sigma}({ x}) \in R \rightarrow R$ is a piecewise-nonlinear function,
\begin{align} \label{oth}
f_{\sigma}({ x})= \left\{ \begin{array} {ccc} {f^I({ x}), } & \quad \mbox{if} \quad {g({x}) \leq \sigma,} \\ {f^{II}({ x}), } &  \quad \mbox{otherwise}, \end{array} \right.
\end{align}
where $f^I ({x}) \in R \rightarrow R$, $f^{II} ({x}) \in R \rightarrow R$, $g({x}) \in R \rightarrow R$ and $\sigma$ is a scalar.
For the sake of simplicity and without loss of generality, here  we assume  that the true system state $x$ is a scalar quantity. The more general case that the state is $n$-dimensional is discussed in an example that we present successively.

Our goal is to identify the unknown parameter $\sigma$ from knowledge of the temporal evolution of $x^k$.
To this aim, we try to model the dynamics of the true system by,
\begin{equation}
{y}^{k+1}=f_{{\sigma}'}(\tilde{{y}}^k), \label{dmodel}
\end{equation}
where ${\sigma}'$ is an estimate of the unknown true coefficient $\sigma$, and $\tilde{{y}}$ is defined as,
\begin{equation}
\tilde{{y}}^k=\epsilon {{y}}^k + (1-\epsilon) { x}^k, \label{eps}
\end{equation}
$\epsilon \in R$. Note that the model is coupled to the true system through $x^k$ in (\ref{eps}). We proceed under the assumption that the value of $\epsilon$ in (\ref{eps}) is such that when $\sigma'=\sigma$, the synchronized solution $y=x$, is stable. Our strategy (to be specified in what follows) seeks to  synchronize the model with the true systems, by dynamically adjusting $\sigma'$ to match the unknown true value of $\sigma$.
We now introduce the following \emph{potential} \cite{SOTT},
\begin{equation}
\Psi=[x-y]^2,   \label{P}
\end{equation}
$\Psi \geq 0$ by definition; $\Psi = 0$ when $y=x$, that is, when the true system and the model system are synchronized. Our goal is to minimize the potential (i.e., to achieve synchronization between the two systems) by dynamically adjusting the estimate $\sigma'$.
Thus, we propose to adaptively evolve $\sigma'$ in time, according to the following gradient descent relation,
\begin{equation}
{\sigma'}^{k+1}-{\sigma'}^{k}=-\eta \frac{\partial \Psi}{\partial \sigma'}=2 \eta [x^k-y^k] \frac{\partial {{y}^k}}{\partial \sigma'} \equiv 2 \eta [x^k-y^k] p^k , \label{dad}
\end{equation}
$\eta>0$. We note  that the term ${p}^k \equiv {\partial {{y}^k}}/{\partial \sigma'}$ appears in Eq. (\ref{dad}); therefore, we seek to find 
a recurrence equation that describes the evolution of $p^k$ in time.
We note that ($\ref{dmodel}$) can be rewritten as,
\begin{equation}
{y}^{k+1}= f^I({\tilde{y}}^k) {\mathcal{H}}({\sigma'}-g({\tilde y}^k))+f^{II}({\tilde y}^k) [1-{\mathcal{H}}({\sigma'}-g({\tilde y}^k))], \label{dheav}
\end{equation}
where $\mathcal{H}$ is the Heaviside step function, $\mathcal{H}(x)=1$ if $x \geq 0$, $0$ otherwise. This allows us to write,
\begin{subequations}\label{dr3}
\begin{align}
& p^{k+1}=  a^k p^k +b^k,  \quad \mbox{with,} \label{dr} \\
& a^k=  \epsilon\{ Df^I({\tilde{y}^k}) {\mathcal{H}}(\sigma'-g({\tilde y}^k))+Df^{II}({\tilde y}^k) [1-{\mathcal{H}}(\sigma'-g({\tilde y}^k))]+[f^{II}({\tilde y}^k)-f^I({\tilde{y}}^k)] {\delta}(\sigma'-g({\tilde y}^k)) Dg({\tilde y}^k)  \}, \label{dra}\\
& b^k=  [f^I({\tilde{y}}^k) -f^{II}({\tilde y}^k)] {\delta}(\sigma'-g({\tilde y}^k))  \label{drb},
\end{align}
\end{subequations}
where $\delta(\cdot)$ is the Dirac delta function. 
 (\ref{dr3}a) is an \emph{auxiliary} difference equation that completes the formulation of our adaptive strategy. In fact, by iterating the set of equations (\ref{dmodel},\ref{dad},\ref{dr3}) and by using knowledge of the time-evolution of the true system state $x_k$,  we 
obtain a time evolving estimate $\sigma'$ of the unknown quantity $\sigma$.  Thus, the adaptive strategy 
is fully described by the set of equations  (\ref{dmodel},\ref{dad},\ref{dr3}).

Figure 1 shows the results of a numerical experiment in which Eq. (\ref{oth}) is the tent map equation, $f^I(x)=\mu x$, $f^{II}(x)=(1-\mu) x$, $\mu=1.4$, $\sigma=0.6$. For these values of the parameters, the tent map dynamics is chaotic. Moreover, we choose $g(x)=x$, $h(x)=x$, $\epsilon=0.55$.
One of the definitions of the Dirac delta function is the following,
\begin{equation}
\delta(u)=\lim_{\varepsilon \rightarrow 0^+} \delta_{\varepsilon}(u)= 2 \lim_{\varepsilon \rightarrow 0^+} {\varepsilon}^{-1} \max(1-|u/\varepsilon|,0)
\end{equation}
In our numerical simulations in Fig. 1, we have approximated $\delta(u)$ by $\delta_{\varepsilon}(u)$, with $\varepsilon=0.1$.
We initialize both the true and the model systems from uniformly distributed random initial conditions between $0$ and $1$, and $\sigma'$ is evolved from an initial estimate that is far away from the true value of $\sigma$, i.e., $\sigma'(0)=0.31$.
As can be seen from Fig. 1, our adaptive strategy is successful in identifying $\sigma$ and in achieving synchronization between the true and the model systems.

\begin{figure}[h]
\centering
\includegraphics[width=0.9\textwidth]{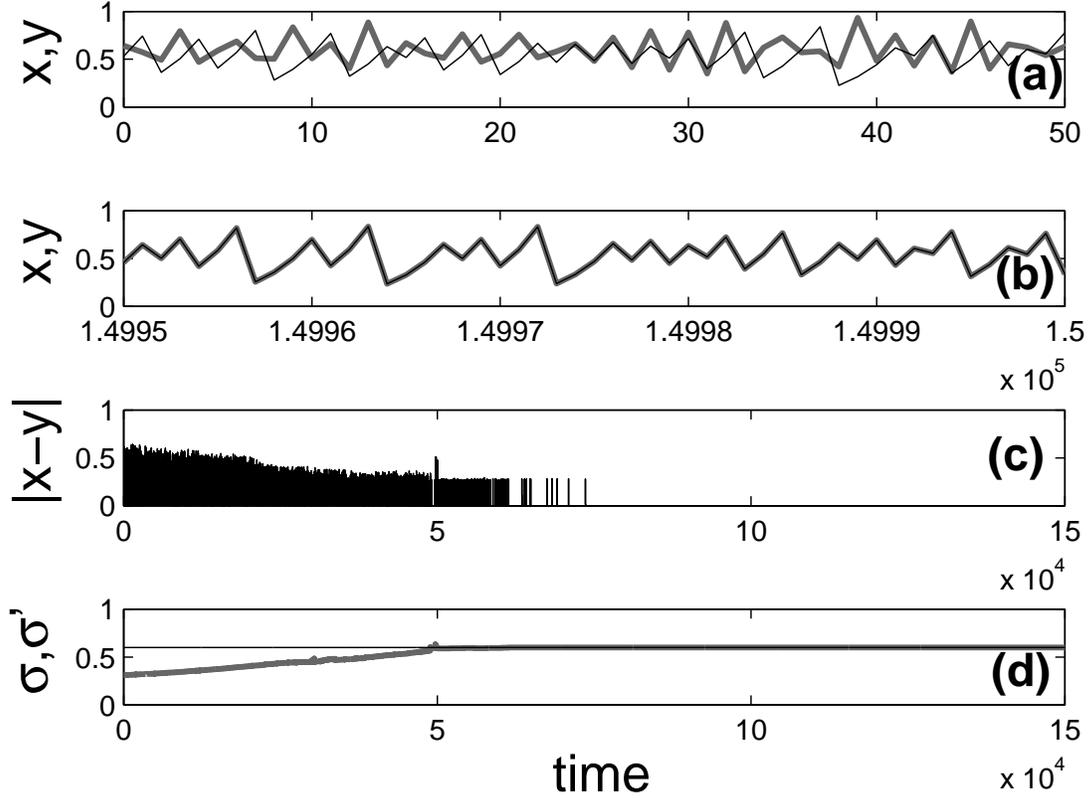}
\caption{(a) and (b) show the time evolutions of $x(k)$ (in black, thin line) and $y(k)$ (in grey, thick line) respectively at the beginning ($k \in [0,50]$) and at the end ($k \in [149950,150000]$) of the run (in (b) the two curves are superposed). Plot (c) shows $|x(k)-y(k)|$ versus $k \in [0,1.5 \times 10^5]$. As can be seen, perfect synchronization is achieved for $k$ approximately greater than $7.5 \times 10^4$. (d) shows the time evolution of $\sigma'(k)$ (in grey), which converges to the true value of $\sigma=0.6$ (the $\sigma$-th ordinate line is plotted as a black solid line in (d)). $f^I(x)=\mu x$, $f^{II}(x)=\mu (1- x)$,  $\mu=1.4$, $\sigma=0.6$, $g(x)=x$, $h(x)=x$, $\epsilon=0.55$, $\eta=10^{-5}$, $\varepsilon=0.1$.}
\end{figure}

Our approach can be extended to the case of continuous-time piecewise dynamical systems, and whose state is in general $n$-dimensional. Moreover, our strategy can be generalized to identify more than one discontinuity point. Though we do not report here a formal formulation of our strategy to encompass all these possible variations, in what follows we present a numerical example, which addresses the case of a continuous-time $n$-dimensional piecewise system, $n=3$. 
Namely, we consider that the true system dynamics is described by the Chua equation, ${\bf{x}}(t)=[x_1(t),x_2(t),x_3(t)]^T$, $\dot{\bf{x}}(t)=F_{\sigma}({\bf{x}}(t))$,
\begin{equation} \label{Chua}
F_{\sigma}({\bf{x}})=\left[\begin{array}{c}  \alpha(x_2(t)-x_1(t)-\phi_{\sigma}(x_1(t)), \\
  x_1(t)-x_2(t)+x_3(t), \\
  -\beta x_2(t), \end{array} \right],
\end{equation}
where the piecewise scalar function $\phi_\sigma(x)$ is defined as,
\begin{align} \label{phi}
\phi_{\sigma}({x})= \left\{ \begin{array} {ccc} {m_1 x +\sigma (m_0-m_1), } & \quad  {{x} \geq \sigma,} \\ m_0{ x}, &  \quad {|x| < \sigma,} \\ {m_1 x -\sigma (m_0-m_1), } & \quad  {{x} \leq -\sigma.} \end{array} \right.
\end{align}
For our choice of $\alpha=15.6$ and $\beta=25.58$, $m_0=-8/7$, $m_1=-5/7$, the Chua system (\ref{Chua},\ref{phi}) displays chaos (the emergence of a chaotic `double scroll' attractor has been observed both in numerical simulations and in experiments \cite{Matsu}).
In the case of (\ref{phi}), the function $\phi_{\sigma}(x)$ has two discontinuity points, i.e., at $\pm \sigma$. In the general case in which $\nu$ discontinuity points are present, $\nu$ independent gradient descent relations can be derived for each of the points and simultaneously integrated (together with other $\nu$ corresponding auxiliary equations, analogous to Eq. (\ref{dr3})) in order to identify them all.  Yet, in the case of the Chua system (\ref{Chua}), the problem can be simply formulated in only one unknown $\sigma$ (i.e., we rely on the fact that the two discontinuity points are symmetrical with respect to zero).

\begin{figure}[h]
\centering
\includegraphics[width=0.9\textwidth]{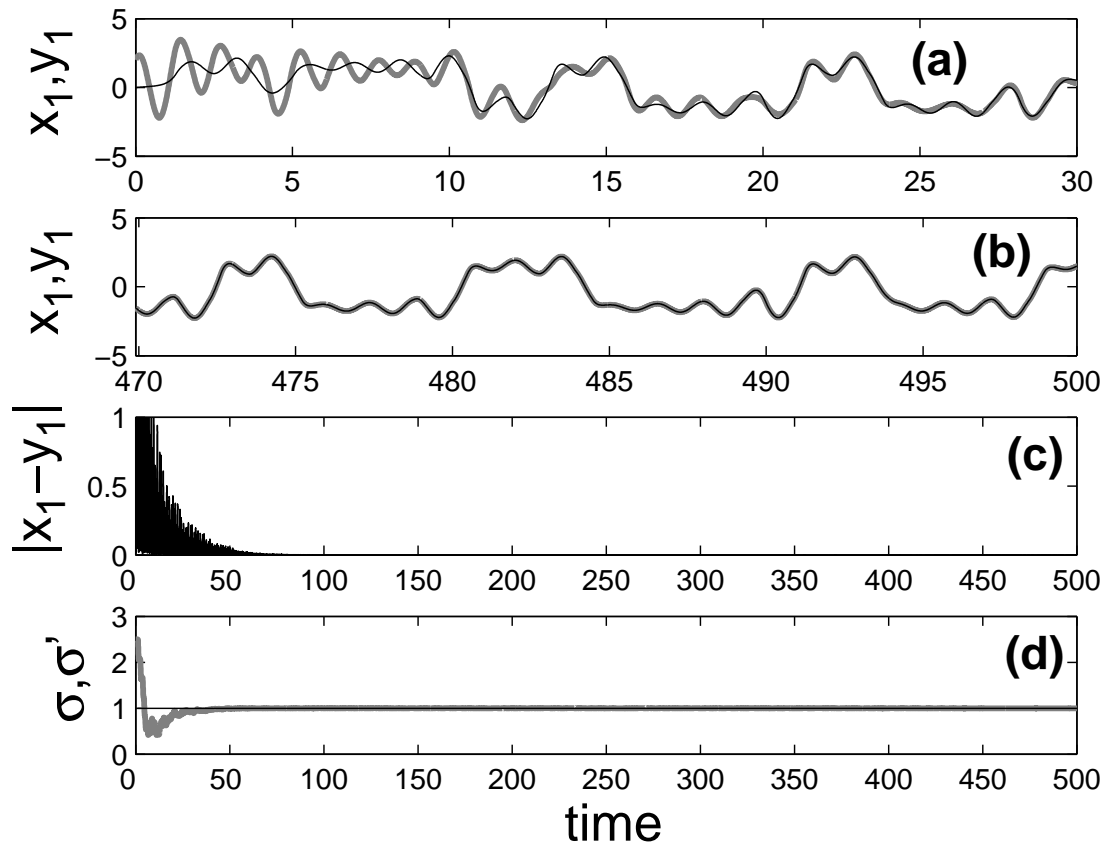}
\caption{(a) and (b) show the time evolutions of $x_1(t)$ (in black, thin line) and $y_1(t)$ (in grey, thick line) respectively at the beginning ($t \in [0,30]$) and at the end ($t \in [470,500]$) of the run (in (b) the two curves are superposed). Plot (c) shows $|x_1(t)-y_1(t)|$ versus $t \in [0,500]$. As can be seen, synchronization is achieved for $t$ approximately greater than $10^2$. (d) shows the time evolution of $\sigma'(t)$ (in grey), which converges to the true value of $\sigma=1$ (the $\sigma$-th ordinate line is plotted as a black thin solid line in (d)). The true system obeys Eq. (\ref{Chua}),  $(\alpha,\beta,m_0,m_1)=(15.6,25.58,-8/7,-5/7)$, with $\sigma=1$ and the model system obeys Eq. (\ref{cmodel}), where the estimate $\sigma'$ evolves according to Eqs. (\ref{csigma}, \ref{cr3}), $h({\bf x})=x_1$, $\Gamma=[\gamma,0,0]^T$, $\gamma=15$, $\zeta=1$. The initial conditions for the true and the model systems are randomly chosen points from the Chua double scroll attractor, $\sigma'$ is evolved from an initial value that is far away from the true value of $\sigma$, i.e., $\sigma'(0)=2.5$, $q_1(0)=0$.}
\end{figure}

We assume to model the true system by ${\bf{y}}(t)=[y_1(t),y_2(t),t_3(t)]^T$,  $\dot{\bf{y}}(t)=F_{\sigma'}({\bf{y}}(t))$, where ${\sigma}'$ is an estimate of the unknown true parameter $\sigma$. We design an adaptive strategy to dynamically adjust $\sigma'$ to match the unknown value of $\sigma$. To this aim, we perform a one way diffusive coupling from the true system to the model, as follows,
\begin{equation}
\dot{{{\bf y}}}(t)=F_{\sigma'}({\bf{y}}(t))+ \Gamma[h({\bf{x}})-h({\bf{y}})] , \label{cmodel}
\end{equation}
where $h({\bf{x}})$ is in general an $m\leq n$ vector of $m$ observable scalar quantities that are assumed to be known functions of the system state $ {\bf{x}}(t)$. $\Gamma$ is an $n \times m$ constant coupling matrix.
In what follows, we assume for simplicity that $h$ is a scalar function ($m=1$),  $h({\bf x})=x_1$, $\Gamma=[\gamma,0,0]^T$ and we proceed under the assumption that the value of $\gamma$ is such that when $\sigma'=\sigma$, the synchronized solution, ${\bf y}= {\bf x}$, is stable. We introduce the following potential,
\begin{equation}
\Psi=[h({\bf x})-h({\bf y})]^2.   \label{P}
\end{equation}
Again, $\Psi \geq 0$ by definition, and $\Psi = 0$ if $h({\bf y})=h({\bf x})$, that is, when the true system and the model system are synchronized. Thus, we seek to minimize the potential by adaptively evolving $\sigma'$ according to the following gradient descent relation,
\begin{equation}
\dot{\sigma'}=-\zeta \frac{\partial \Psi}{\partial \sigma'}=2 \zeta [h({\bf{x}})-h({\bf{y}})] Dh({\bf{y}}) \frac{\partial {\bf{y}}}{\partial \sigma'} \equiv 2 \zeta [h({\bf{x}})-h({\bf{y}})] Dh({\bf{y}}) {\bf{q}}, \label{csigma}
\end{equation}
$\zeta>0$. Note that for our choice of $h$, $Dh=[1, 0, 0]$, yielding $Dh {\bf{q}}=q_1$, where $q_1$ is the first component of the vector ${\bf q}$. Therefore, we seek a differential equation that describes how ${q_1} \equiv {\partial {{y}_1}}/{\partial \sigma'}$ evolves in time.
We note that (\ref{phi}) can be rewritten as,
\begin{equation}
\phi_{\sigma}({y_1})=m_0 y_1+(m_1-m_0) [\mathcal{H}(-\sigma'-y_1) (y_1+\sigma')+ \mathcal{H}(y_1-\sigma')(y_1-\sigma')]. \label{phisc}
\end{equation}
Then, from Eqs. (\ref{cmodel},\ref{phisc}), we obtain the following differential equation for $q_1$,
\begin{subequations}\label{cr3}
\begin{align}
& \dot{q}_1(t)=  c(t) q_1(t) +d(t),  \quad \mbox{where,}\\
&c(t)=  -\gamma-\alpha \{1+ m_0+ (m_1-m_0) [\mathcal{H}(-\sigma'-y_1)+\mathcal{H}(y_1-\sigma')]\}, \label{cra}\\
&d(t)=  -\alpha (m1-m0) [\mathcal{H}(-\sigma'-y_1)-\mathcal{H}(y_1-\sigma')].  \label{crb}
\end{align}
\end{subequations}
Our adaptive strategy is then fully described by Eqs. (\ref{cmodel}, \ref{csigma}, \ref{cr3}). In order to test the strategy, we run a numerical experiment (shown in Fig. 2), in which we integrate Eqs. (\ref{Chua},\ref{cmodel}, \ref{csigma}, \ref{cr3}). We initialize both the true and the model systems from uniformly distributed random initial conditions on the Chua chaotic attractor, and $\sigma'$ is evolved from an initial value that is far away from the true value of $\sigma=1$, i.e., $\sigma'(0)=2.5$, $q_1(0)=0$.
As can be seen, our adaptive strategy is successful in identifying $\sigma$ and in achieving synchronization between the true and the model systems.

Hereafter, we consider a completely different identification problem, 
in which the dynamics of the system to be identified is described by a set of delay differential equations,
\begin{equation}
\dot{\bf x}(t)=\Phi\big({\bf x}(t),{\bf x}(t-\tau)\big),
\end{equation}
where ${\bf x}=(x_1,x_2,...,x_n)^T$, $\Phi \in R^n \rightarrow R^n$, $\tau$ is a time delay.

We try to model the dynamics of the true system by  $\dot{\bf y}(t)=\Phi({{{\bf y}(t)}},{{{\bf y}(t-\tau')}})$, where $\tau'$ is an estimate of the unknown true coefficient $\tau$. Our goal is to evolve $\tau'$ in time to match the unknown true $\tau$-value and in so doing, to achieve synchronization between the model and the true systems. Therefore, we perform a one way diffusive coupling from the true system to the model, as follows,
\begin{equation}
\dot{{{\bf y}}}(t)=\Phi({{{\bf y}(t)}},{{{\bf y}(t-\tau'(t))}})+ \Gamma[h({\bf{x}}(t))-h({\bf{y}}(t))] , \label{modelt}
\end{equation}
where  $h({\bf{x}})$ and $\Gamma$ are the same as defined before. 
In what follows, we assume for simplicity that $h$ is a scalar function, i.e., $m=1$.

We now introduce the potential $\Psi$, defined in (\ref{P}) (similar considerations apply as in the previous cases) and
we propose to minimize $\Psi$ 
by making $\tau'$ converge onto the true value $\tau$, through the following gradient descent relation,
\begin{equation}
\dot{ \tau'}=-\beta \frac{\partial \Psi}{\partial \tau'}=2 \beta [h({\bf y}(t))-h({\bf x}(t))] Dh \frac{\partial {\bf{y}}'}{\partial \tau'}=2 \beta [h({\bf y}'(t))-h({\bf y}(t))] Dh {\bf r} , \label{ad1t}
\end{equation}
$\beta>0$.
 We are now interested in how the $n$-vector ${\bf r}= {\partial {\bf{y}}'}/{\partial \tau'}$ evolves in time. We note that the following equation describes the evolution of ${\bf r}(t)$,
 \begin{equation}
 \dot{{\bf r}}(t)=\big[\frac{\partial \phi}{\partial {\bf y}'(t)}-\Gamma Dh \big] {\bf r}(t) - \frac{\partial \phi}{\partial {\bf y'}(t-\tau)}{\dot{\bf y}'(t-\tau)}. \label{ad2t}
 \end{equation}
Therefore our adaptive strategy 
is fully described by the set of equations  (\ref{modelt},\ref{ad1t},\ref{ad2t}).

We now present a numerical experiment testing the above strategy for a case in which the unknown true system is described by the Mackey-Glass equation, for which $n=1$, ${\bf y}(t) \equiv y(t)$, $\phi(y(t),y(t-\tau))=-b y(t)+a y(t-\tau)/(1+y(t-\tau)^{10})$, $\Gamma$ coincides with the scalar coupling $\gamma$. We choose $a=0.2$, $b=0.1$, $\tau=23$, $h(y)=y$. For these values of the parameters, the dynamics of the Mackey-Glass system is chaotic.

 Both the true system and the model system are initialized from  uniformly distributed random initial conditions on the chaotic attractor. We have preliminarily found that for $\tau'=\tau=23$, the two systems synchronize for $\gamma>\gamma^*=0.07$. Thus, for our numerical experiment, we choose $\gamma=0.1$.
Moreover, we take $\beta=1$. The results of our numerical experiment are shown in Fig. 3. The experiment consists of two parts. For $t \leq 10^4$, $\tau$ is kept constant and equal to $23$, while $\tau'$ is initialized from a value that is far away from the true value of $\tau=23$, i.e., $\tau'(0)=15$. As can be seen, after a transient, $t \gtsimeq 0.25 \times 10^4$, $\tau'$ converges to the true value of $\tau=23$ and the synchronization error $|x-y|$ decays to zero. For $10^4<t \leq 2 \times 10^4$, $\tau$ becomes a function of time, i.e., $\tau(t)=23+3 \sin(2 \pi 10^{-4} t)$, while the adaptive strategy (\ref{modelt},\ref{ad1t},\ref{ad2t}) is kept running. As can be seen, for $t> 10^4$, $\tau'(t)$ tracks quite well the time evolution of $\tau(t)$ and approximate synchronization between the model and the true system is attained.   

\begin{figure}[h]
\centering
\includegraphics[width=0.9\textwidth]{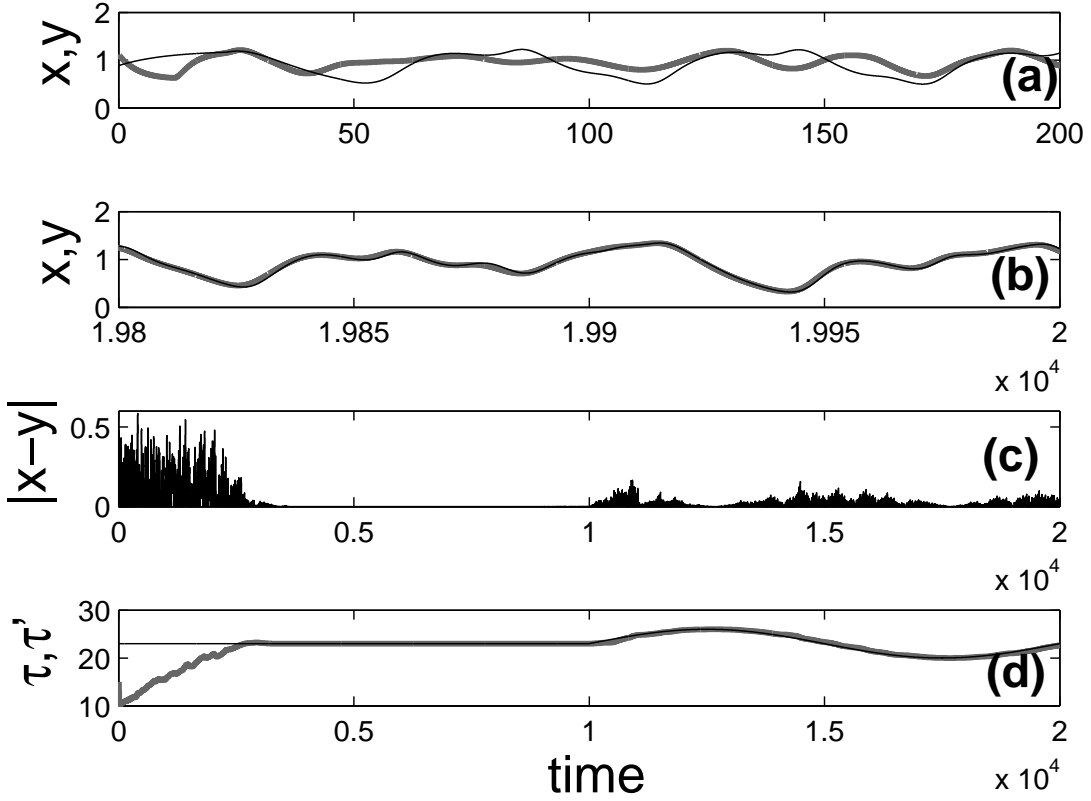}
\caption{(a) and (b) show the time evolutions of $x(t)$ (in black, thin line) and $y(t)$ (in grey, thick line) respectively at the beginning ($t \in [0,2 \times 10^2]$) and at the end ($t \in [19800,2 \times 10^4]$) of the run. Plot (c) shows $|x(t)-y(t)|$ versus $t \in [0,2 \times 10^4]$. (d) shows the time evolution of $\tau'(t)$ (in grey, thick line), compared to the time evolution of $\tau(t)$ (in black, thin line). The experiment consists of two parts. For $t \leq 10^4$, $\tau$ is kept constant and equal to $23$; for $10^4<t \leq 2 \times 10^4$, $\tau$ becomes a function of time, i.e., $\tau(t)=23+3 \sin(2 \pi 10^{-4} t)$.  In the first part, after an initial transient, the estimate $\tau'$ converges to the true value of $\tau=23$; in the second part, $\tau'$ tracks the time evolution of $\tau(t)$.   $\phi(y(t),y(t-\tau))=-b y(t)+a y(t-\tau)/(1+y(t-\tau)^{10})$, $a=0.2$, $b=0.1$; $h(y)=y$, $\gamma=0.1$, $\beta=1$. In our experiment, the true system and the model system are initialized from uniformly distributed random initial conditions on the chaotic attractor, $\tau'(0)=15$.}
\end{figure}

Synchronization has been shown to be a convenient tool to identifying the dynamics of unknown systems. Different techniques have been proposed, see e.g., \cite{Abarbanel,Yu:Parlitz,IDTOUT}. Yet, the problems of identification of  delays and discontinuity points that commonly characterize the dynamics of real systems have not received adequate attention in the literature. In this paper, we have proposed a general methodology that, based on a simple gradient descent technique (MIT rule), can be successfully employed to address both these problems. We have also shown the usefulness of our strategy in the case that the unknown parameters of the true system to be estimated slowly drift in time (here, by `slowly' we mean that they evolve on a time scale which is much longer than that on which a typical chaotic oscillation occurs).

We successfully apply our methodology to solve problems as different as the identification of delays and the identification of discontinuity points of unknown dynamical systems. This suggests that our approach can provide a simple and flexible paradigm for the resolution of a 
variety of diverse identification problems. Furthermore, by making use of the properties of synchronization of chaos, the adaptive strategy can be used to extract several unknowns from knowledge of the chaotic state-evolution of the system of interest.

\end{document}